\documentclass[a4paper]{jpconf}
\usepackage{graphicx}
\begin{document}
\title{Canonical quantization of anisotropic Bianchi I cosmology
from scalar vector tensor Brans Dicke gravity}
\author{Hossein Ghaffarnejad}
\address{Faculty of Physics, Semnan University, Semnan, IRAN, 35131-19111}
\ead{ hghafarnejad@semnan.ac.ir}
\begin{abstract}
We applied a generalized scalar-vector-tensor Brans Dicke gravity
model to study canonical quantization of an anisotropic Bianchi I
cosmological model. Regarding an anisotropic Harmonic Oscillator
potential we show that the corresponding Wheeler de Witt wave
functional of the system is described by Hermit polynomials.
 We obtained
a quantization condition on the ADM mass of the cosmological
system which raises versus the quantum numbers of the Hermit
polynomials. Our calculations show that the inflationary expansion
of the universe can be originate from the big bang with  no naked
singularity due to the uncertainty principle.
\end{abstract}
\section{Introduction}

The standard cosmology has a great success in explaining the
observations of the cosmic microwave background radiation (CMBR)
temperature \cite{LL,HD,DM,We}. This model is based on the
validity of the cosmological principle (the spatial homogeneity
and isotropy but not in the time direction) and the Einstein`s
general relativity explain most large-scale observations with
unprecedented accuracy. However, several directional anomalies
have been reported in various large-scale observations (see
\cite{Ghafhoda} and references therein).  They have no place in
standard cosmology and are not being studied.
  In fact origin of these anomalies do not
understood and so there are two different proposals to understand
them as follows: a) Perhaps they are originated from cosmological
effects which should be described via alternative gravity theories
instead of the Einstein`s general theory of relativity. b) Other
possibility which arises these directional anomalies can be
systematic errors or contaminations of measuring instruments and
etc., which should be exclude from the future data analysis. In
the latter case one usually accept validity of the standard
cosmological $\Lambda CDM$ model while in the former proposal one
use an alternative gravity model instead of the Einstein`s general
theory of relativity. Zhao and Santos, provided full review about
these proposals in ref. \cite{Zh} where the directional anomalies
predict a preferred axis $"Axis~of~Evil"$ in large scale of the
Universe.
 However, the general covariance principal leads
us to believe that these anomalies have cosmological origin and it
can be described by some alternative gravity models where the
cosmological principals should be violated (see (\cite{Ghafhoda})
and reference therein). To describe the above mentioned anomalies
the anisotropic Bianchi cosmological models are applicable
\cite{Cam, Cam2, Cam3} for anisotropic cosmological constant and
dark energy. As an alternative gravity model we consider
scalar-vector-tensor gravity model \cite{Ghaf,Ghaf2} which is made
from generalization of the well known Jordan-Brans-Dicke scalar
tensor gravity \cite{BDD} by transforming the background metric as
$g_{\mu\nu}\to g_{\mu\nu}+2N_{\mu}N_{\nu}$. $N_{\mu}$ is dynamical
four vector field which can be called as four velocity of a
preferred reference frame. Several classical and quantum
applications of this model are studied previously for FLRW
cosmology  which are addressed in references of the work
\cite{Ghaf3}. As an application of the gravity model for
anisotropic cosmological models we applied it to study Bianchi I
classical cosmology in ref. \cite{Ghafhoda}  where the
corresponding Freedmann equations read to anisotropic inflationary
expanding universe and they satisfied observational data
successfully. In the present work we want to resolve the naked
singularity of the anisotropic Bianchi I cosmological model by
applying the canonical quantization approach. To do so we use
self-interaction anisotropic three dimensional Harmonic Oscillator
potential to calculate Hamiltonian operator of the system. Then we
obtain Wheeler de Witt wave functional of the anisotropic Bianchi
I background metric versus the Hermit polynomials. Outlook of this
work is to present a quantization condition on the energy density
of the system with no naked singularity where the system is stable
at high energy quantum level. Organization of the paper is as
follows. \\ In section 2, we introduce the scalar-vector-tensor
gravity model \cite{Ghaf,Ghaf2} under consideration. In section 3,
we use the Bianchi I type of the background metric to obtain exact
form of Hamiltonian density. In section 4 we solve Wheeler de Witt
wave equation of the system and obtain a quantization condition on
the energy density of the Bianchi I quantum cosmological system.
The obtained Wheeler de Witt wave solution is described versus the
quantum anisotropic harmonic Oscillator (Hermit polynomials) for
three dimensional anisotropic Oscillator potential. In section 5
we discuss about the quantization of the ADM energy of the system.
In section 6 we discuss about the big bang naked singularity of
the expanding universe which how can it is removed in the quantum
perspective of the system. Section 7 denotes to concluding remark.

 \section{The gravity model}

Let us start with the Brans-Dicke scalar-vector tensor gravity
\cite{Ghaf, Ghaf2}
\begin{equation}\label{acf}
 I=\frac{1
}{16\pi}\int d^4x\sqrt{-g}\left\{\phi R-\frac{\omega}{
    \phi}
g^{\mu\nu}\nabla_{\mu} \phi\nabla_{\nu}\phi
\right\}$$$$+\frac{1}{16\pi}\int
d^4x\sqrt{-g}\{\zeta(x^{\nu})(g^{\mu\nu}N_{\mu}N_{\nu}+1)+2\phi
F_{\mu\nu}F^{\mu\nu}+U(\phi,N_{\mu})$$$$-\phi N_\mu
N^{\nu}(2F^{\mu\lambda}\Omega_{\nu\lambda}+
F^{\mu\lambda}F_{\nu\lambda}+\Omega^{\mu\lambda}\Omega_{\nu\lambda}
-
2R^{\mu}_{\nu}+\frac{2\omega}{\phi^2}\nabla^{\mu}\phi\nabla_{\nu}\phi)\},
\end{equation}
  where $g$ is absolute value of determinant of the
metric tensor $g_{\mu \nu}$ with signature  as (-,+,+,+), $\phi$
is the Brans-Dicke scalar field and $\omega$ is the Brans-Dicke
adjustable coupling constant. The tensor fields $F_{\mu\nu}$ and
$\Omega_{\mu\nu}$ are defined versus the time like vector field
$N_{\mu}$ as follows.
\begin{equation}\label{fo}
F_{\mu\nu}=2(\nabla_{\mu}N_{\nu}-\nabla_{\nu}N_{\mu}),~~~~~~~\Omega_{\mu\nu}=2(\nabla_{\mu}N_{\nu}+\nabla_{\nu}N_{\mu})
\end{equation}
 The above action is written in units
$c=G=\hbar=1$ and the undetermined Lagrange multiplier
$\zeta(x^{\nu})$ controls $N_{\mu}$ to be an unit time-like vector
field. $\phi$ describes inverse of variable Newton's gravitational
coupling parameter and its dimension is $(lenght)^{-2}$ in units
$c=G=\hbar=1$. Present limits of dimensionless BD parameter
$\omega$ based on time-delay experiments \cite{Will, Will2,Gaz,
Rea} requires $\omega\geq4\times10^{4}.$ Varying the above action
functional with respect to $\zeta$ we obtain time-like condition
on the vector field $N_{\mu}$ as follows.
\begin{equation}\label{vect}g_{\mu\nu}N^{\mu}N^{\nu}=-1.\end{equation}
In the next section we apply the above mentioned action functional
to study the anisotropic Bianchi I cosmological model.
\section{Bianchi I quantum cosmology}
Spatially homogenous but anisotropic dynamical flat universe is
defined by the Bianchi I metric which from point of view of
 free falling (comoving) observer is defined by the following line element \cite{AK11}.
 \begin{equation}\label{met} ds^2=-dt^2+e^{2a(t)}\{e^{-4b(t)}dx^2+e^{2b(t)}(dy^2+dz^2)\}\end{equation}
where ${x,y,z}$ are cartesian coordinates of the comoving observer
and $t$ is cosmic time.  In the above metric equation we assume
that the spatial part has cylindrical symmetry for which
$e^{a(t)}$ is an global isotropic scale factor and $b(t)$
represents a deviation from the isotropy.
 Substituting (\ref{met}) into the equation (\ref{vect}) we obtain
 \begin{equation}\label{vectt} N_{\mu}(t)=\left(%
\begin{array}{c}
  N_t \\
  N_x \\
  N_y \\
  N_z \\
\end{array}%
\right)=\left(%
\begin{array}{c}
  \cosh\alpha \\
  e^{a-2b}\sinh\alpha\cos\beta \\
  e^{a+b}\sinh\alpha\sin\beta\cos\gamma \\
  e^{a+b}\sinh\alpha\sin\beta\sin\gamma \\
\end{array}%
\right)\end{equation} where $(\alpha,\beta,\gamma)$ are angular
constant parameters of the vector field $N_{\mu}$ which makes as
fixed its direction at the 4D anisotropic space time (\ref{met}).
  Substituting (\ref{met}) and (\ref{vectt}) into the action functional (\ref{acf}) with some simple calculations, one can
  show that the action functional read
 \begin{equation}\label{action}I=\frac{1}{16\pi}\int dxdydz\int dt e^{3a}\bigg[-\omega\frac{\dot{\phi}^2}{\phi}-6\phi\dot{a}^2+6\phi\dot{b}^2
 -U(a,b,\phi)
 \bigg]\end{equation} where dot $\dot{}$ denotes to `cosmic` time derivative $\frac{d}{dt}$
 and we used the ansatz $\alpha=0$
 to eliminate frictional terms $\ddot{a}$ and $\ddot{b}$ in the  action functional (\ref{action}) (see ref.\cite{Ghafhoda} for more detail).
 In general, if an action
functional contains time derivative of velocity of the dynamical
fields then there
 will be some frictional forces
  which cause that the extremum point of the action functional does not fixed. The latter
kind of
 dynamical systems are not closed and so stable. They behave usually as chaotic dynamical systems. Hence we should eliminate
 these acceleration terms of the dynamical fields to fix extremum points of the
 system. Usually one obtained an effective action functional by
 integrating by part and removing divergence-less counterpart of
 the action functional.
  Here we can eliminate
the frictional terms $\ddot{a}$ and $\ddot{b}$ without to use an
effective action functional instead of the action functional
(\ref{action}) just by setting $\alpha=0$. This restrict us to
choose a particular direction for the time-like dynamical vector
field (\ref{vectt}) where the lagrangian of the system has not
friction terms $\ddot{a}$ and $\ddot{b}$.\\ When we study
canonical quantization of a mini-super-space quantum cosmology,
then there the local coordinates do not have an important role but
the dynamical fields themselves play an important role. Hence it
will be useful to apply a conformal frame with the following
conformal time $\tau(t)$ in what follows.
\begin{equation}\label{conf}dt=F(t)d\tau,~~~~F=\frac{e^{3a}\phi}{16\pi}\end{equation}
where $F$ is called as lapse (red shift) function in the ADM
formalism of the decomposition of the background metric.
Substituting (\ref{conf}) and
\begin{equation}\label{psi}\phi=\frac{e^{Z(\tau)}}{G}\end{equation}
into the action functional (\ref{action}) we obtain $I=\int
dxdydz\int\mathcal{L}d\tau$ where $G$ is Newton`s coupling
constant and  $\mathcal{L}$ is the Lagrangian density of the
system from point of view of the conformal frame.
\begin{equation}\label{Lag}\mathcal{L}=-\omega Z^{\prime2}-6a^{\prime2}+6b^{\prime2}-V(a,b,Z)\end{equation}
where $\prime$ denotes to derivative with respect to the conformal
time $\tau$ and
\begin{equation}\label{pot}V=\frac{F^2U}{\phi}\end{equation} is a
suitable super-potential. Calculating canonical momenta of the
fields $(a,b,Z)$ as
\begin{equation}\label{mom}\Pi_a=\frac{\partial\mathcal{L}}{\partial a^{\prime}},~~~\Pi_b=\frac{\partial\mathcal{L}}{\partial b^{\prime}},~~~\Pi_Z=
\frac{\partial\mathcal{L}}{\partial Z^{\prime}}\end{equation} and
applying definition of the Hamiltonian density
\begin{equation}\label{Ham}\mathcal{H}=a^{\prime}\Pi_{a}+b^{\prime}\Pi_b+Z^{\prime}\Pi_Z-\mathcal{L}\end{equation}
one can infer
\begin{equation}\label{Hamm}\mathcal{H}=-\frac{\Pi_a^2}{24}+\frac{\Pi_b^2}{24}-\frac{\Pi_Z^2}{4\omega}+V(a,b,Z)\end{equation}
where we see that signature of the de Witt superspace metric is
Lorentzian form $(-,+,-).$ In the next section we see that the
canonical momentum operator of the anisotropic counterpart of the
metric field $b$ behaves as a time-evolution parameter of the
system in the canonical quantum cosmology of the minisuperspace de
Witt metric. However we obtain now Wheeler de Witt probability
wave solution of the cosmological system under a quantization
condition on the ADM energy of the system.
\section{Canonical quantization}
To study quantum stability of the Bianchi I cosmology we should
first fix the potential $V(a,b,Z)$ where we choose anisotropic
harmonic Oscillator potential defined on the mini-super-space de
Witt metric as follows.
\begin{equation}\label{harmpot}
V(a,b,Z)= \frac{1}{2}K_a a^2-\frac{1}{2}K_bb^2+\frac{1}{2}K_ZZ^2.
\end{equation}
 Substituting the Dirac`s canonical
quantization condition for the momentum operators as
\begin{equation}\label{dirac}\hat{\Pi}_a=-i\frac{\delta}{\delta a},~~~\hat{\Pi}_b=-i\frac{\delta}{\delta b},~~~\hat{\Pi}_Z
=-i\frac{\delta}{\delta Z},~~~\end{equation} and the potential
form (\ref{harmpot}) into the Hamiltonian density (\ref{Hamm}) we
obtain the corresponding Wheeler de Witt wave equation
$\hat{\mathcal{H}}\Psi(a,b,X)=0$ as follows.
\begin{equation}\label{dewitt}\bigg\{\frac{1}{24}\frac{\delta^2}{\delta a^2}-\frac{1}{24}\frac{\delta^2}{\delta b^2}+
\frac{1}{4\omega}\frac{\delta^2}{\delta Z^2}-
\frac{K_a}{2}a^2+\frac{K_b}{2}b^2-\frac{K_Z}{2}Z^2\bigg\}\Psi=0.\end{equation}
Negativity sign of the differential operator
$\frac{\delta^2}{\delta b^2}$ in the above equation shows that the
anisotropy field $Y$ can behave as the time evolution parameter in
the three dimensional de Witt superspace metric. In other word the
Kinetic terms in the above differential equation is similar to a
Klein Gordon operator defined on the mini-super-space de Witt
metric. To solve the equation (\ref{dewitt}) one can apply the
standard method of separation of variables as follows. We assume
the Wheeler de Witt wave solution to be separable as
$\Psi(a,b,Z)=A(a)B(b)C(Z)$ and substitute it into the equation
(\ref{dewitt}), then we can obtain the following differential
equations for the fields $A(a), B(b), C(Z)$ respectively as
follows.
\begin{equation}\label{A}\bigg[\frac{1}{24}\frac{\delta^2}{\delta a^2}+\epsilon_a-\frac{K_a}{2}a^2\bigg]A(a)=0\end{equation}

\begin{equation}\label{B}\bigg[\frac{1}{24}\frac{\delta^2}{\delta b^2}+\epsilon_b-\frac{K_b}{2}b^2\bigg]B(b)=0\end{equation}

\begin{equation}\label{C}\bigg[\frac{1}{4\omega}\frac{\delta^2}{\delta Z^2}+\epsilon_Z-\frac{K_Z}{2}Z^2\bigg]C(Z)=0\end{equation}
where the constants of separation of variables $\epsilon_{a,b,Z}$
satisfy the following relation.
\begin{equation}\label{epsilon}\epsilon_a+\epsilon_Z=\epsilon_b.\end{equation}
One can show that the equations (\ref{A}), (\ref{B}) and (\ref{C})
 can be described by the well known Hermit polynomials if we set the following quantization conditions on the parameters $\epsilon_{X,Y,Z}.$
 \begin{equation}\label{quantum}\epsilon_a=\bigg(N_a+\frac{1}{2}\bigg)\sqrt{\frac{K_a}{12}},~~~\epsilon_b=\bigg(N_b+\frac{1}{2}\bigg)
 \sqrt{\frac{K_b}{12}}
 ,~~~\epsilon_Z=\bigg(N_Z+\frac{1}{2}\bigg)\sqrt{\frac{K_Z}{2\omega}}\end{equation}
in which   $N_{\{a,b,Z\}}$ can take quantized values
$0,1,2,3,\cdots .$ In the latter case we will
 obtain normalized form of the solutions $A(a), B(b)$ and $C(Z)$
 as  follows.
 \begin{equation}\label{A(x)} A_{N_a}(a)=\frac{H_{N_a}(\lambda_a a)e^{-\lambda_a a^2/2}}{\sqrt{2^{N_a}N_a!\sqrt{\pi}}},~~~\lambda_a=(12K_X)^{\frac{1}{4}}\end{equation}

\begin{equation}\label{B(y)} B_{N_b}(b)=\frac{H_{N_b}(\lambda_b b)e^{-\lambda_b b^2/2}}{\sqrt{2^{N_b}N_b!\sqrt{\pi}}},~
~~\lambda_b=(12K_b)^{\frac{1}{4}}\end{equation}

\begin{equation}\label{C(z)} C_{N_Z}(Z)=\frac{H_{N_Z}(\lambda_Z Z)e^{-\lambda_Z Z^2/2}}{\sqrt{2^{N_Z}N_Z!\sqrt{\pi}}},~~~\lambda_Z=(2\omega K_Z)^{\frac{1}{4}}\end{equation}
in which  $H_{N_a}, H_{N_b}, H_{N_Z}$ are Hermit polynomials.
Multiplying the above solutions we obtain $\Psi_{N_aN_bN_Z}=
A_{N_a}(a)B_{N_b}(b)C_{N_Z}(Z)$ which in fact describes quantum
fluctuations of the metric field of the Bianchi I cosmology
defined by (\ref{met}) when the cosmological system is in quantum
 state $(N_a,N_b,N_Z)$ with corresponding eigne energy $(\epsilon_{N_a}, \epsilon_{N_b},\epsilon_{N_Z})$. Now we can show that general solution of
the Wheeler de Witt wave equation (\ref{dewitt}) by expanding it
versus the eigne functionals such that
\begin{equation}\label{wheeler}\Psi(X,Y,X)=\Sigma_{N_a=0}^\infty\Sigma_{N_b=0}^{\infty}\Sigma_{N_Z=0}^\infty
D_{N_aN_bN_Z}\Psi_{N_aN_bN_Z}(a,b,Z)\end{equation} where the
coefficient $D_{N_XN_YN_Z}$ describes probability of the quantum
Bianchi I cosmology which takes the eigne state
$\Psi_{N_XN_YN_Z}.$ This can be determined by regarding the
initial condition of the system and orthogonal condition on the
Hermit polynomials.  However no one does not know about initial
condition of "physical cosmology" but we can predicts some
physical statements about our obtained solutions in what follows.
Substituting (\ref{quantum}) into the condition (\ref{epsilon}) we
can obtain allowable eigne states of the system as follows.
\begin{equation}\label{allow}\bigg(N_a+\frac{1}{2}\bigg)\sqrt{\frac{K_a}{K_b}}+\bigg(N_Z+\frac{1}{2}\bigg)\sqrt{\frac{6}{\omega}}\sqrt{\frac{K_Z}{K_b}}
=\bigg(N_b+\frac{1}{2}\bigg).\end{equation}
\section{ADM energy and mass}
In the previous section we assume the ADM mass of the cosmological
system has a zero value. If it is not permissible then we should
solve the extended Wheeler de Witt wave equation as
$\hat{\mathcal{H}}\Psi=\mathcal{M}\Psi$ in which $\mathcal{M}$ is
called to be the ADM mass of the system. In the latter case the
equation (\ref{epsilon}) should be extended to the following form.
\begin{equation}\label{mass}\mathcal{M}=\epsilon_b-(\epsilon_a+\epsilon_Z)\end{equation}
which by substituting the quantization conditions (\ref{quantum})
we obtain a quantization condition on the ADM mass of the Bianchi
I cosmological model such that
\begin{equation}\label{massquant}\mathcal{M}_{N_aN_bN_Z}=
\bigg(N_b+\frac{1}{2}\bigg)\sqrt{\frac{K_b}{12}}-\bigg(N_a+\frac{1}{2}\bigg)\sqrt{\frac{K_a}{12}}-\bigg(N_Z+\frac{1}{2}\bigg)\sqrt{\frac{K_Z}{2\omega}}\end{equation}
In fact ADM energy is a special way to define the energy in
general relativity, which is only applicable to some special
geometries of spacetime that asymptotically approach a
well-defined metric tensor at infinity. If the background metric
approaches to Minkowski space asymptotically then the Neother`s
theorem implies that the ADM energy or mass should be invariant
because of time independence of the Minkowski metric. According to
general relativity, the conservation law for the total energy does
not hold in more general. For instance for time-dependent
background metrics it will be violated. For example, it is
completely violated in physical cosmology. In fact cosmic
inflation in particular is able to produce energy or mass from
"nothing" because the vacuum energy density is roughly constant,
but the volume of the Universe grows exponentially. Here we can
modeled the violation of the ADM energy in the anisotropic Bianchi
I inflationary cosmology by quantization approach. Here we show
that the raising the ADM energy in the Bianchi I cosmology can be
described by quantum fluctuations of the mass parameters of the
fields $(a,b,Z)$ which we called with $K_a,K_b,K_Z$ respectively.
\section{Big Bang naked singularity}
The background metric (\ref{met}) shows that the naked singularity
lies on the particular hypersurface $(a,b)\to(-\infty,-\infty)$ at
the classical cosmological feature where the big bang originated.
Applying $(a,b)\to(-\infty,-\infty)$ one can infer that  for naked
singularity  we have $A(-\infty)=0=B(-\infty).$ This reads
\begin{equation}\label{lim}\lim_{(a,b)\to(-\infty,-\infty)}\Psi(a,b,Z)=0.\end{equation}
Physical interpretation of the above result can be described as
follows: There is a zero probability where the big bang originates
from a naked singularity. In other words the big bang originates
from an anisotropic quantum Oscillation of the quantum matter
fields $(a,b,Z)$ at small scales of the space time. This happened
because of the uncertainty relations between the fields and the
corresponding canonical momenta as $\Delta
X_i\Delta\Pi_{X_i}\equiv1$ where $X_i=(a,b,Z).$

\section{Concluding remarks}
In this paper we choose scalar vector tensor Brans Dicke gravity
to study anisotropic Bianchi I cosmology in the canonical
quantization approach. We solved Wheeler de Witt wave equation and
obtained its solutions versus the Hermit polynomials for an
anisotropic harmonic Oscillator potential. We obtained a
quantization condition on the ADM boundary energy of the system.
Mathematical calculations predict that the inflation of the
universe can be originated from a big bang state without a naked
singularity. Furthermore anisotropic counterpart of the metric
field can play as a time evolution parameter on the
mini-super-space because of Lorentzian signature of the de Witt
metric. As a future work one can study backreaction of the Hawking
thermal radiation of the quantum matter (the vector and the Brans
Dicke scalar) fields on the solutions of the Wheeler de Witt
equations and so the corresponding eigne energies (quantized ADM
energy) of the system which does not considered here.
 
\end{document}